\documentclass[12pt]{article}
\usepackage{amsfonts, latexsym}
\usepackage{array}
\usepackage{amsmath}
\usepackage{amsthm}
\newtheorem{proposition}{Proposition}

\begin{document}

\title{\Huge Towards an Effective Field Theory of QED}
\author{J. Kijowski \\ Center for Theor. Phys., Polish Academy of Sciences \\
                   al. Lotnik\'ow 32/46, 02 - 668 Warsaw, Poland \\
                   \ \\
    G. Rudolph, M. Rudolph \\
                   Institut f\"ur Theoretische Physik, Univ. Leipzig \\
                   Augustusplatz 10/11, 04109 Leipzig, Germany  }
\maketitle

\begin{abstract}
A procedure for reducing the functional integral of QED to an integral over
bosonic gauge invariant fields is presented.
Next, a certain averaging method for this integral, giving a
tractable effective quantum field theory, is proposed. Finally, the
current--current propagator and the chiral anomaly are calculated within
this new formulation. These results are part of our programme of analyzing
gauge theories with fermions in terms of local gauge invariants.
\end{abstract}

\section{Introduction}
\label{Introduction}

This paper is part of our programme of analyzing gauge theories in terms
of physical observables (i.e. gauge invariants).  For applications of
this programme to non-Abelian Higgs models we refer to
\cite{1}. In recent years we have applied it to theories of
gauge fields interacting with fermionic matter fields, see \cite{KR1} -- \cite{KRR2}. 
In \cite{KR1} we have proved that the
classical Dirac-Maxwell system can be formulated in a spin-rotation
covariant way in terms of gauge invariant quantities. In \cite{KRR1} we
have shown that similar constructions work on the level of the (formal)
functional integral of QED, where fermion fields are treated as
anticommuting (Berezian) quantities, and in \cite{KRR2} we have applied
our procedure to the 2--dimensional Schwinger model. As a result we
obtained a functional integral completely reformulated in terms of local
gauge invariant quantities, which differs essentially from the effective
functional integral obtained via the Faddeev-Popov procedure \cite{FP}.

The present paper is a continuation of \cite{KRR1}. It turns out that our
general procedure leads to a complicated, singular functional integral
kernel. In order to make this model tractable, we propose a certain
averaging procedure leading to an effective quantum field theory. This
theory is characterized by a certain number of parameters, which have to
be fixed by comparison with experimental data. As an application we
discuss -- for the massless case -- the current--current propagator and
the chiral anomaly within this formulation. (In principle, the massive
case can be also dealt with, using an expansion in the mass parameter,
but this problem will be not addressed in this paper.) A number of
interesting phenomena and results comes out: A bosonization rule,
similar to that in the Schwinger model appears naturally. Moreover, we
get a dynamical mass generation leading to a massive spin--1 field.
Physical quantities like the current--current propagator and the chiral anomaly
are given as expectation values with respect to an effective non-local measure.
This measure can be analyzed in terms of a power series expansion in the 
coupling constant, which, however, is
completely different from the ordinary perturbation expansion. This is
due to the fact that the above mentioned mass itself contains the bare
coupling constant. Therefore, the formulae obtained suggest that
automatically some resumation of the ordinary perturbation series has
taken place. It is also remarkable that our formulation leads to a
completely new approach to calculate the chiral (Adler-Bardeen) anomaly.
In lowest order this quantity can be calculated analytically. Adjusting
part of our free parameters yields the standard Adler-Bardeen anomaly
with the correct coefficient.

We stress, that our approach circumvents any gauge fixing and,
therefore, also the Gribov problem \cite{Grib}. It leads naturally to
bosonization and can be viewed as a general construction scheme for
effective quantum field theories. Due to the above remarks, it seems to
be appropriate for the study of non--perturbative aspects. We also mention
that a similar construction is possible for lattice models within the
Hamiltonian framework. In this context we have discussed the charge
superselection structure of QED \cite{6}.

Our method applies also to Yang-Mills theories, see \cite{KRR3}, where
the functional integral of one-flavour chromodynamics is reduced to an
integral over purely bosonic invariants.

\section{QED in Terms of Local Gauge Invariants}
\label{GaugeInvariantForm}

The functional integral of QED is given by
\begin{eqnarray}
\label{FctInt}
Z & = & \int \prod {\rm d}A\,  {\rm d}\psi\,  {\rm d}\psi^* \,
        e^{i \, \int {\rm d}^4 x \, {\cal L}[A,\psi,\psi^* ]}, \\
{\cal L}
  & = & {\cal L}_{gauge} + {\cal L}_{mat} \nonumber\\
  & = & - \frac{1}{4} F_{\mu \nu} F^{\mu \nu} - m \psi^{a*}
        \, \beta_{ab} \,  \psi^b - Im \left \{\psi^{a*} \, \beta_{ab} \,
        {\left(\gamma^{\mu}\right)^b}_c \, D_{\mu}\psi^c \right \},
\end{eqnarray}
where $F_{\mu \nu} = {\partial}_{\mu} A_{\nu} - {\partial}_{\nu}
A_{\mu}$ and $D_{\mu}\psi^a = {\partial}_{\mu}\psi^a + ie A_{\mu}
\psi^a$ are the electromagnetic field strength and the covariant
derivative, respectively. Here $a,b, ... = 1,2,\dot 1,\dot 2$ denote
bispinor indices and $\mu,\nu, ... = 0,1,2,3$ spacetime indices,
$\beta_{ab}$ denotes the Hermitean metric in bispinor space and
${\left(\gamma^{\mu}\right)^b}_c$ are the Dirac matrices. The
anticommuting components of the bispinor field $\psi^a$, which can be
represented by a pair of Weyl spinors $\psi^a = \left( \begin{array}{c}
\phi^K \\  \varphi_{\dot K} \end{array} \right)$,
generate a Grassmann--algebra of pointwise real dimension 8. The
representation used for these quantities can be found in \cite{KR1}
and \cite{KRR1}.

In \cite{KRR1} we have proposed a procedure which reduces the functional
integral (\ref{FctInt}) to an integral over gauge invariants. It 
is based upon the following ideas: First one has to analyse the
algebra of Grassmann--algebra--valued gauge invariants, which can be
built from the gauge potential $A_{\mu}$ and the anticommuting matter
fields $\psi^a$. Typically, there occurs a number of identities between
the invariants which, in general, cannot be solved on the algebraic level.
In particular, one finds a relation which expresses the Lagrangian, or
the Lagrangian multiplied by some non--vanishing element of the above
algebra, in terms of invariants. In a next step one has to implement
this relation under the functional integral and to reduce the original
functional integral measure to a measure in terms of invariants. For
that purpose we make use of the following notion of the
$\delta$--distribution on superspace (see \cite{5} and \cite{Ber2})
\begin{equation}
\delta(u - U)
 = \int {\rm d} \xi \, e^{2 \pi {\rm i} \xi (u - U)} =
   \sum_{n=0}^{\infty} \frac{(-1)^n}{n!} \delta^{(n)}(u) U^n.
\end{equation}
Here $u$ is a c--number variable and $U$ an element of the
Grassmann--algebra built from matter fields $\psi$ and $\psi^*$.
>From this definition we get immediately
\begin{equation}
\label{delt1}
1 \equiv \int {\rm d}u \, \delta(u-U).
\end{equation}
Thus, by inserting identites of the form (\ref{delt1}) under the
functional integral, we introduce for each Grassmann--algebra--valued
gauge invariant a c--number variable, which we call c--number mate.
These mates are by definition gauge invariant. This way we are able to
solve the above-mentioned relations between invariants, leading to a
theory reformulated in terms of gauge invariant fields. For details we
refer to \cite{KRR1} and \cite{KRR3}.

For QED we start with the following Grassmann--algebra--valued gauge
invariants:
\begin{eqnarray} H & := & {\varphi}_K^* {\phi}^K , \label{Inv1} \\
B_{\mu} & := & {\rm Im} \left\{ H^* \left( {\varphi}_K^* D_{\mu} {\phi}^K +
{\phi}^K  D_{\mu} \varphi_K^* \right) \right\}, \label{Inv2} \\ J^{\mu} & := &
\psi^{a*} \, \beta_{ab} \, (\gamma^\mu)^{b}_{\ c} \, \psi^c \label{Inv3} \\
J_5^{\mu} & := & \psi^{a*} \, \beta_{ab} \, {(\gamma^{\mu})^b}_c \,
{(\gamma_5)^c}_d \, \psi^d. \label{Inv4}
\end{eqnarray}
Here $H$ is a complex scalar field, whereas $B_{\mu}$ is a real-valued
covector field. $J^{\mu}$ and $J_5^{\mu}$ denote the vector and
axial--vector current, respectively.

We denote the corresponding c-number mates by $h$, $b_{\mu}$, $j^{\mu}$
and $j_5^{\mu}$, and put $v_{\mu} := \frac{b_{\mu}}{2e \, |h|^2}$ as
well as $h = |h| e^{i \, \alpha}$. It was shown in \cite{KRR1} that our
procedure yields the following functional integral:
\begin{equation}
\label{InvFI1}
Z = \int \prod_x \{ {\rm d}v_{\mu} \, {\rm d}j^{\mu} \, {\rm d}j^{\mu}_5 \,
    {\rm d}|h|^2 \, {\rm d}\alpha \, K[j^{\mu},j^{\mu}_5,|h|^2] \} \,
    e^{{\rm i} \int {\rm d}^4 x \, {\cal L}[v_{\mu},j^{\mu},j^{\mu}_5,|h|,\alpha]},
\end{equation}
where
\begin{eqnarray}
\label{InvIntKernel}
K[j^{\mu},j_5^{\mu},|h|^2]
 & = &   \frac{1}{16 \, \pi} \, \Big\{
         \frac{\delta^2}{\delta j^{\mu} \delta j_{\mu}}
         \frac{\delta^2}{\delta j^{\nu} \delta j_{\nu}}
      +  2 \frac{\delta^2}{\delta j^{\mu} \delta j_{\mu}}
         \frac{\delta^2}{\delta j_5^{\nu} \delta j^5_{\nu}}
      -  4 \frac{\delta^2}{\delta j^{\mu} \delta j^5_{\mu}}
         \frac{\delta^2}{\delta j^{\nu} \delta j^5_{\nu}} \nonumber \\
 & &  +  \frac{\delta^2}{\delta j_5^{\mu} \delta j^5_{\mu}}
         \frac{\delta^2}{\delta j_5^{\nu} \delta j^5_{\nu}}
      +  \frac{1}{16} \, \frac{\delta^4}{\delta |h|^4}
      +  \frac{1}{8 \, |h|} \, \frac{\delta^3}{\delta |h|^3}
      -  \frac{1}{16 \, |h|^2} \, \frac{\delta^2}{\delta |h|^2} \nonumber \\
 & &  +  \frac{1}{16 \, |h|^3} \, \frac{\delta}{\delta |h|}
      -  \frac{1}{2} \, \frac{\delta^2}{\delta |h|^2}
         \frac{\delta^2}{\delta j^{\mu} \delta j_{\mu}}
      -  \frac{1}{2 \, |h|} \, \frac{\delta}{\delta |h|}
         \frac{\delta^2}{\delta j^{\mu} \delta j_{\mu}} \nonumber \\
 & &  +  \frac{1}{2} \, \frac{\delta^2}{\delta |h|^2}
         \frac{\delta^2}{\delta j_5^{\mu} \delta j^5_{\mu}}
      +  \frac{1}{2 \, |h|} \, \frac{\delta}{\delta |h|}
         \frac{\delta^2}{\delta j_5^{\mu} \delta j^5_{\mu}}
      \Big\} \,
      \delta(j^{\mu})  \delta(j_5^{\mu}) \delta(|h|^2).
\end{eqnarray}
and
\begin{eqnarray}
\label{InvLag1}
{\cal L}[v_{\mu},j^{\mu},j^{\mu}_5,|h|,\alpha] & = &
- \tfrac{1}{4} (\partial_{[\mu}\,v_{\nu]})^2
- e \, j^{\mu} v_{\mu}
+ \tfrac{1}{2} \, j_5^{\mu} (\partial_{\mu} \alpha) \\
& & + \tfrac{1}{8 \, |h|^2} \, \epsilon^{\alpha \beta \mu \gamma} \,
  j_{\alpha} j^5_{\beta} (\partial_{\mu} j_{\gamma})
  - 2 \, m \, |h| \cos \alpha. \nonumber
\end{eqnarray}

\section{Effective Bosonized QED}
\label{SEffectiveBosonizedForm}

Observe that the integral kernel $K[j^{\mu},j^{\mu}_5,|h|^2]$ has the
form $K = {\cal D} \left\{\delta(j^{\mu}) \delta(j^{\mu}_5)
\delta(|h|^2) \right\}$, where ${\cal D}$ is a differential operator
containing functional derivatives with respect to $j^{\mu}, j_5^{\mu}$
and $|h|$, multiplied by singular coefficients. A priori, this
expression does not make sense. In order to regularize it, we replace it
by a Gaussian measure with three free parameters, $\alpha_j$, $\alpha_{j^5}$
and $\alpha_h$, which have to be fixed by physical requirements later. This
way we get:
\begin{equation}
\label{InvFI2}
Z = {\cal N} \, \int \prod_x \{ {\rm d}v_{\mu} \, {\rm d}j^{\mu} \,
    {\rm d}j^{\mu}_5 \, {\rm d}|h|^2 \, {\rm d}\alpha \} \,
    e^{{\rm i} \int {\rm d}^4 x \, {\cal L}[v_{\mu},j^{\mu},j^{\mu}_5,|h|,\alpha]},
\end{equation}
where now
\begin{eqnarray}
\label{InvLag2}
{\cal L}[v_{\mu},j^{\mu},j^{\mu}_5,|h|,\alpha] & = &
   - \tfrac{1}{4} (\partial_{[\mu}\,v_{\nu]})^2
   - e \, j^{\mu} v_{\mu}
   + \tfrac{1}{2} \, j_5^{\mu} (\partial_{\mu} \alpha)
   + \tfrac{1}{8 \, |h|^2} \, \epsilon^{\alpha \beta \mu \gamma} \,
     j_{\alpha} j^5_{\beta} (\partial_{\mu} j_{\gamma}) \nonumber \\
& &  - 2 m \, |h| \cos \alpha - \tfrac{1}{2 \alpha_j} \, j^{\mu} j_{\mu}
   - \tfrac{1}{2 \alpha_{j^5}} \, j^{\mu}_5 j_{\mu}^5
   - \tfrac{1}{2 \alpha_h} \, |h|^4 \  .
\end{eqnarray}

It is interesting to note that this regularization can be achieved by a
technical trick similar to that used in the Faddeev--Popov procedure:
One can average the singular kernel (\ref{InvIntKernel}) with a
functional depending on three auxiliary fields corresponding to the
variables $j^{\mu}, j_5^{\mu}$ and $|h|$. It was shown in \cite{R} that
the requirement to obtain the above Gaussian measure after this
averaging determines this functional uniquely.

We see that the Lagrangian (\ref{InvLag2}) does not contain derivatives
of the chiral current $j^5_{\mu}$, i.e. $j^5_{\mu}$ enters the theory as
a non--dynamical field. Thus, we can carry out the simple Gaussian
integral over the chiral current, which yields
\begin{equation}
\label{InvFI3}
Z
 = {\cal N} \, \int \prod_x \big\{ {\rm d} v_{\mu} \, {\rm d} j^{\mu} \,
   {\rm d} |h|^2 \, {\rm d} \alpha \big\} \,
   e^{{\rm i} \int {\rm d}^4 x \,
     {\cal L}[v_{\mu},j^{\mu},|h|,\alpha]} \  ,
\end{equation}
where the effective Lagrangian has taken the form:
\begin{eqnarray}
\label{InvLag3}
{\cal L}[v_{\mu},j^{\mu},|h|,\alpha]
 & = & - \tfrac{1}{4} (\partial_{[\mu}\,v_{\nu]})^2
       - e \, j^{\mu} v_{\mu}
       + \tfrac{\alpha_{j^5}}{16 \, |h|^2} \,
         \epsilon^{\alpha \mu \beta \gamma} \,
         (\partial_{\mu} \alpha) j_{\alpha} (\partial_{\beta} j_{\gamma})
         \nonumber \\
 &   & + \tfrac{\alpha_{j^5}}{128 \, |h|^4} \,
         \epsilon^{\alpha \mu \beta \gamma} \,
         \epsilon_{\delta \mu \rho \sigma} \,
         j_{\alpha} (\partial_{\beta} j_{\gamma})
         j^{\delta} (\partial^{\rho} j^{\sigma}) \\
 &   & - \tfrac{1}{2 \alpha_j} \, j^{\mu} j_{\mu}
       + \tfrac{\alpha_{j^5}}{8} \, (\partial^{\mu} \alpha) (\partial_{\mu}
       \alpha) - \tfrac{1}{2 \alpha_h} \, |h|^4
       - 2 m \, |h| \cos \alpha. \nonumber
\end{eqnarray}

Comparing this Lagrangian with (\ref{InvLag2}) we observe that $j^5_{\mu}$
has been replaced by the gradient of $\alpha$. More precisely, we have the
remarkable relation
\begin{equation}
\label{QEDBos}
\partial_{\mu} \alpha = \frac{2}{\alpha_{j^5}} \, j^5_{\mu}\  .
\end{equation}
This is the 4-dimensional analog of the bosonization rule in the
2-dimensional Schwinger model, see \cite{Schw}. (In \cite{KRR2} we have
obtained this rule for the Schwinger model using our approach.)

Observe that the field $|h|^2$ enters (\ref{InvLag3}) 
in a non--dynamical way, too.
Thus, in principle, one should integrate it out. This, however, can not
be done explicitly. In a first approximation, $|h|^2$ could be replaced
by a constant. We will come to that point later.

Finally, let us write down the generating functional integral of our
effective theory (\ref{InvFI3}), (\ref{InvLag3}):
\begin{equation}
\label{GFI}
Z[\zeta^{\mu}, \xi^{\mu}, \eta^{\mu}]
 = {\cal N} \, \int \prod_x \{ {\rm d}v_{\mu} \, {\rm d}j^{\mu} \,
   {\rm d}|h|^2 \, {\rm d}\alpha \} \,
   e^{{\rm i} \int {\rm d}^4 x \,
     {\cal L}[\zeta^{\mu}, \xi^{\mu}, \eta^{\mu};
     v_{\mu},j^{\mu},|h|,\alpha]}
     \  ,
\end{equation}
with
\begin{eqnarray}
\label{GLag}
\lefteqn{ {\cal L}[\zeta^{\mu}, \xi^{\mu}, \eta^{\mu};
 v_{\mu},j^{\mu},|h|,\alpha] }
\nonumber \\
 & \hspace*{1cm} = & - \tfrac{1}{4} (\partial_{[\mu}\,v_{\nu]})^2
       - e \, j^{\mu} v_{\mu}
       + \tfrac{\alpha_{j^5}}{16 \, |h|^2} \,
         \epsilon^{\alpha \mu \beta \gamma} \,
         (\partial_{\mu} \alpha) j_{\alpha} (\partial_{\beta} j_{\gamma})
         \nonumber \\
 &   & + \tfrac{\alpha_{j^5}}{128 \, |h|^4} \,
         \epsilon^{\alpha \mu \beta \gamma} \,
         \epsilon_{\delta \mu \rho \sigma} \,
         j_{\alpha} (\partial_{\beta} j_{\gamma})
         j^{\delta} (\partial^{\rho} j^{\sigma})
       - \tfrac{1}{2 \alpha_j} \, j^{\mu} j_{\mu}
       - \tfrac{1}{2 \alpha_h} \, |h|^4 \nonumber \\
 &   & + \tfrac{\alpha_{j^5}}{8} \, (\partial^{\mu} \alpha)
        (\partial_{\mu} \alpha) - 2 m \, |h| \cos \alpha + \zeta^{\mu}
       (\partial_{\mu} \alpha) + \xi^{\mu} j_{\mu} + \eta^{\mu} v_{\mu}
       \  , \hspace*{1cm}
\end{eqnarray}
where $\zeta^{\mu}$, $\xi^{\mu}$ and $\eta^{\mu}$ denote the source
currents for $(\partial^{\mu} \alpha)$, $j^{\mu}$ and $v^{\mu}$,
respectively.

\section{The Current--Current Propagator}
\label{CurrentCurrentPropagator}

In this Section we want to calculate the current--current propagator
$<0|T j^{\mu}(y_1) \, j^{\nu}(y_2)|0>$ using the effective
theory obtained in the last Section. We restrict ourselves
to the massless case, i.e. we put $m = 0$ in the generating functional
integral (\ref{GFI}).

The current--current propagator is given by
\begin{eqnarray}
\lefteqn{ <0|T j^{\mu}(y_1) \, j^{\nu}(y_2)|0> } \nonumber \\
 & \hspace*{1cm} = &
       \frac{1}{Z[0,0,0]} \, \frac{1}{{\rm i}^2} \,
       \frac{\delta}{\delta \xi_{\mu}(y_1)} \,
       \frac{\delta}{\delta \xi_{\nu}(y_2)} \,
       Z[\zeta^{\mu}, \xi^{\mu}, \eta^{\mu}]
       \Big|_{\zeta^{\mu},\xi^{\mu}, \eta^{\mu} = 0} \nonumber \\
 & \hspace*{1cm} = & \frac{1}{Z[0,0,0]} \, \frac{1}{{\rm i}^2} \,
       \frac{\delta}{\delta \xi_{\mu}(y_1)} \,
       \frac{\delta}{\delta \xi_{\nu}(y_2)} \,
       Z[0, \xi^{\mu}, \eta^{\mu}]
       \Big|_{\xi^{\mu}, \eta^{\mu} = 0}. \label{CC1}
\end{eqnarray}
To handle the non--linear (self--interaction) term
$$\tfrac{\alpha_{j^5}}{128 \, |h|^4} \, \epsilon^{\alpha \mu \beta \gamma}
\, \epsilon_{\delta \mu \rho \sigma} \, j_{\alpha} (\partial_{\beta}
j_{\gamma})  j^{\delta} (\partial^{\rho} j^{\sigma})$$ occuring in
(\ref{GLag}) we introduce the vector field
\begin{equation}
\label{CC2}
k^{\mu}[j_{\mu},|h|]
 := \tfrac{1}{|h|^2} \, \epsilon^{\alpha \mu \beta \gamma} \,
    j_{\alpha} (\partial_{\beta} j_{\gamma}) \  ,
\end{equation}
and decompose it into its longitudinal and transversal parts:
\begin{equation}
\label{CC3}
k^{\mu}[j_{\mu},|h|] = \partial^{\mu} k[j_{\mu},|h|] +
k^{\mu \bot}[j_{\mu},|h|] \  ,
\end{equation}
where $\partial_{\mu} k^{\mu \bot}[j_{\mu},|h|] = 0$.
This yields
\begin{eqnarray}
\label{CC5}
\lefteqn{{\cal L}[\zeta^{\mu}, \xi^{\mu}, \eta^{\mu};
        v_{\mu},j^{\mu},|h|,\alpha] }
          \nonumber \\
 & \hspace*{1cm} = & - \tfrac{1}{4} \, (\partial_{[\mu} v_{\nu]})^2
       - e \, j^{\mu} v_{\mu}
       + \tfrac{\alpha_{j^5}}{16} \,
         (\partial_{\mu} \alpha) \, (\partial^{\mu} k[j_{\mu},|h|])
         + \tfrac{\alpha_{j^5}}{128} \,(\partial^{\mu} k[j_{\mu},|h|])
         (\partial_{\mu} k[j_{\mu},|h|])
          \nonumber \\
 &   & + \tfrac{\alpha_{j^5}}{128} \, k^{\mu \bot}[j_{\mu},|h|] \,
         k_{\mu}^{\bot}[j_{\mu},|h|]
       - \tfrac{1}{2 \alpha_j} \, j^{\mu} j_{\mu}
       - \tfrac{1}{2 \alpha_h} \, |h|^4
       + \tfrac{\alpha_{j^5}}{8} \, (\partial^{\mu} \alpha)
         (\partial_{\mu} \alpha) \nonumber \\
 &   & + \zeta^{\mu}(\partial_{\mu}\alpha) + \xi^{\mu} j_{\mu}
       + \eta^{\mu} v_{\mu}.
\end{eqnarray}
The term $\tfrac{\alpha_{j^5}}{16} \, (\partial_{\mu} \alpha) \,
 k^{\mu \bot}[j_{\mu},|h|]$ vanishes by partial integration.

Transforming
\begin{equation}
\label{CC6b}
\alpha'
 = \alpha + \tfrac{1}{4 \, \alpha_{j^5}} \, k[j_{\mu},|h|],
\end{equation}
we obtain
\begin{eqnarray}
\label{CC7}
\lefteqn{ {\cal L}[0, \xi^{\mu}, \eta^{\mu}; v_{\mu},j^{\mu},|h|,\alpha']}
\nonumber \\
& \hspace*{1cm} = &
    - \tfrac{1}{4} \, (\partial_{[\mu} v_{\nu]})^2
    - e \, j^{\mu} v_{\mu}
    + \tfrac{\alpha_{j^5}}{128} \, k^{\mu \bot}[j_{\mu},|h|] \,
      k_{\mu}^{\bot}[j_{\mu},|h|]
    \nonumber \\
& & - \tfrac{1}{2 \, \alpha_j} \, j^{\mu} j_{\mu}
    - \tfrac{1}{2 \, \alpha_h} |h|^4
    + \tfrac{\alpha_{j^5}}{8} \, (\partial^{\mu} \alpha')
      (\partial_{\mu} \alpha')
    + \xi^{\mu} j_{\mu}
    + \eta^{\mu} v_{\mu}.
\end{eqnarray}
Observe that the transformation (\ref{CC6b}) leaves the integral
measure ${\rm d}\alpha$ invariant. Thus, $\alpha'$ can be integrated out
trivially and we get
\begin{eqnarray}
\lefteqn{ <0|T j^{\mu}(y_1) \, j^{\nu}(y_2)|0> } \nonumber \\
 & \hspace*{1cm} = &
       \frac{1}{Z[0,0,0]} \,
       \int \prod_x \big\{ {\rm d} v_{\mu} \, {\rm d} j^{\mu} \, {\rm d} |h|^2 \big\}
       \nonumber \\
 &   & \hspace*{1.7cm} \times \,
       \frac{1}{{\rm i}^2} \,
       \frac{\delta}{\delta \xi_{\mu}(y_1)} \,
       \frac{\delta}{\delta \xi_{\nu}(y_2)} \,
       e^{{\rm i} \int {\rm d}^4 x \, {\cal L}[0, \xi^{\mu}, \eta^{\mu};
       v_{\mu},j^{\mu},|h|]}
       \Big|_{\xi^{\mu}, \eta^{\mu} = 0}, \hspace*{1cm} \label{CC8}
\end{eqnarray}
where
\begin{eqnarray}
\lefteqn{ {\cal L}[0, \xi^{\mu}, \eta^{\mu}; v_{\mu},j^{\mu},|h|] }
\nonumber \\
& \hspace*{1cm} = &
      - \tfrac{1}{4} \, (\partial_{[\mu} v_{\nu]})^2
      - e \, j^{\mu} v_{\mu}
      + \tfrac{\alpha_{j^5}}{128} \, k^{\mu \bot}[j_{\mu},|h|] \,
        k_{\mu}^{\bot}[j_{\mu},|h|] \nonumber \\
&   & - \tfrac{1}{2 \, \alpha_j} \, j^{\mu} j_{\mu}
      - \tfrac{1}{2 \, \alpha_h} |h|^4
      + \xi^{\mu} j_{\mu}
      + \eta^{\mu} v_{\mu}. \label{CC9}
\end{eqnarray}

The further analysis of the term
$\tfrac{\alpha_{j^5}}{128} \, k^{\mu \bot}[j_{\mu},|h|] \,
   k_{\mu}^{\bot}[j_{\mu},|h|]$ needs some care, due to its non--linear (and
non--local) character.
Expanding the exponential of this term in a series, we obtain
\begin{eqnarray}
\lefteqn{ <0|T j^{\mu}(y_1) \, j^{\nu}(y_2)|0> } \nonumber \\
 & \hspace*{1cm} = &
       \frac{1}{Z[0,0,0]} \,
       \int \prod_x \Big\{ {\rm d} v_{\mu} \, {\rm d} j^{\mu} \,
       {\rm d} |h|^2 \nonumber \\
 &   & \hspace*{1.8cm} \times \,
       \sum_{n=0}^{\infty} \tfrac{1}{n!} \,
           \left( \tfrac{\alpha_{j^5}}{128} \right)^n \,
           \left( k^{\mu \bot}[j_{\mu},|h|] \, k_{\mu}^{\bot}[j_{\mu},|h|]
           \right)^n
       \Big\} \nonumber \\
 &   & \hspace*{1.8cm} \times \,
       \frac{1}{{\rm i}^2} \,
          \frac{\delta}{\delta \xi_{\mu}(y_1)} \,
          \frac{\delta}{\delta \xi_{\nu}(y_2)} \,
       e^{{\rm i} \int {\rm d}^4 x \, {\cal L}_0[0, \xi^{\mu}, \eta^{\mu};
       v_{\mu},j^{\mu},|h|]}
       \Big|_{\xi^{\mu},\eta^{\mu} = 0} \nonumber \\
 & \hspace*{1cm} = & \frac{1}{Z[0,0,0]} \,
       \int \prod_x \Big\{ {\rm d} v_{\mu} \, {\rm d} j^{\mu} \,
       {\rm d} |h|^2 \nonumber \\
 &   & \hspace*{1.8cm} \times \,
       \sum_{n=0}^{\infty} \tfrac{1}{n!} \,
           \left( \tfrac{\alpha_{j^5}}{128} \right)^n \,
           \left( k^{\mu \bot}[\tfrac{\delta}{{\rm i} \,
           \delta \xi^{\mu}},|h|] \,
           k_{\mu}^{\bot}[\tfrac{\delta}{{\rm i} \, \delta \xi^{\mu}},|h|]
           \right)^n
       \Big\} \nonumber \\
 &   & \hspace*{1.8cm} \times \,
       \frac{1}{{\rm i}^2} \,
          \frac{\delta}{\delta \xi_{\mu}(y_1)} \,
          \frac{\delta}{\delta \xi_{\nu}(y_2)} \,
       e^{{\rm i} \int {\rm d}^4 x \, {\cal L}_0[0, \xi^{\mu},
       \eta^{\mu}; v_{\mu},j^{\mu},|h|]}
       \Big|_{\xi^{\mu},\eta^{\mu} = 0}, \nonumber
\end{eqnarray}
where
\begin{eqnarray*}
Z[0,0,0]
 & = & \int \prod_x \Big\{ {\rm d} v_{\mu} \, {\rm d} j^{\mu} \,
 {\rm d} |h|^2 \,
       \sum_{n=0}^{\infty} \tfrac{1}{n!} \,
           \left( \tfrac{\alpha_{j^5}}{128} \right)^n \,
           \left( k^{\mu \bot}[\tfrac{\delta}{{\rm i} \,
           \delta \xi^{\mu}},|h|] \,
           k_{\mu}^{\bot}[\tfrac{\delta}{{\rm i} \, \delta \xi^{\mu}},|h|]
           \right)^n
       \Big\} \nonumber \\
 &   & \times \,
       e^{{\rm i} \int {\rm d}^4 x \,
          {\cal L}_0[0, \xi^{\mu}, \eta^{\mu}; v_{\mu},j^{\mu},|h|]}
       \Big|_{\xi^{\mu},\eta^{\mu} = 0}
\end{eqnarray*}
and
\begin{eqnarray}
\lefteqn{ {\cal L}_0[0, \xi^{\mu}, \eta^{\mu}; v_{\mu},j^{\mu},|h|] }
\nonumber \\
& \hspace*{1cm} = &
      - \tfrac{1}{4} \, (\partial_{[\mu} v_{\nu]})^2
      - e \, j^{\mu} v_{\mu}
      - \tfrac{1}{2 \, \alpha_j} \, j^{\mu} j_{\mu}
      - \tfrac{1}{2 \, \alpha_h} |h|^4
      + \xi^{\mu} j_{\mu}
      + \eta^{\mu} v_{\mu}. \nonumber
\end{eqnarray}
In the last step we used the fact that the integral kernel consists of a
polynomial function in $j^{\mu}$. Thus, $j^{\mu}$ can be replaced by the
corresponding functional derivative with respect to the source current
$\xi^{\mu}$. Now we are left with a Gaussian integral with respect to
$j^{\mu}$, which yields
\begin{eqnarray}
\lefteqn{ <0|T j^{\mu}(y_1) \, j^{\nu}(y_2)|0> } \nonumber \\
 & \hspace*{1cm} = &
       \frac{1}{Z[0,0,0]} \,
       \int \prod_x \Big\{ {\rm d} v_{\mu} \, {\rm d} |h|^2 \,
       \sum_{n=0}^{\infty} \tfrac{1}{n!} \,
           \left( \tfrac{\alpha_{j^5}}{128} \right)^n \,
           \left( k^{\mu \bot}[\tfrac{\delta}{{\rm i} \,
           \delta \xi^{\mu}},|h|] \,
           k_{\mu}^{\bot}[\tfrac{\delta}{{\rm i} \, \delta \xi^{\mu}},|h|]
           \right)^n
       \Big\} \nonumber \\
 &   & \hspace*{1.7cm} \times \,
       \frac{1}{{\rm i}^2} \,
       \frac{\delta}{\delta \xi_{\mu}(y_1)} \,
       \frac{\delta}{\delta \xi_{\nu}(y_2)} \,
       e^{{\rm i} \int {\rm d}^4 x \, {\cal L}_0[0, \xi^{\mu},
       \eta^{\mu};v_{\mu},|h|]}
       \Big|_{\xi^{\mu},\eta^{\mu} = 0} \label{CC10}
\end{eqnarray}
with
\begin{eqnarray*}
Z[0,0,0]
 & = & \int \prod_x \Big\{ {\rm d} v_{\mu} \, {\rm d} |h|^2 \,
       \sum_{n=0}^{\infty} \tfrac{1}{n!} \,
          \left( \tfrac{\alpha_{j^5}}{128} \right)^n \,
          \left( k^{\mu \bot}[\tfrac{\delta}{{\rm i} \,
          \delta \xi^{\mu}},|h|] \,
          k_{\mu}^{\bot}[\tfrac{\delta}{{\rm i} \, \delta \xi^{\mu}},|h|]
       \right)^n
       \Big\} \nonumber \\
 &   & \times \,
       e^{{\rm i} \int {\rm d}^4 x \,
       {\cal L}_0[0,\xi^{\mu},\eta^{\mu};v_{\mu},|h|]}
       \Big|_{\xi^{\mu},\eta^{\mu} = 0}
\end{eqnarray*}
and
\begin{eqnarray}
\lefteqn{ {\cal L}_0[0,\xi^{\mu},\eta^{\mu};v_{\mu},|h|] } \nonumber \\
 & \hspace{1cm} = &
     - \tfrac{1}{4} \, (\partial_{[\mu} v_{\nu]})^2
      + \tfrac{e^2 \, \alpha_j}{2} \, v^{\mu} v_{\mu}
      + \tfrac{\alpha_j}{2} \, \xi^{\mu} \xi_{\mu}
      - \tfrac{1}{2 \, \alpha_h} |h|^4
      - \alpha_j e \, v^{\mu} \xi_{\mu}
      + \eta^{\mu} v_{\mu}. \hspace*{1cm} \label{CC11}
\end{eqnarray}

We remark, that the covector field $v_{\mu}$ has acquired a mass $m_v^2
= \alpha_j e^2$.  Thus, in our effective field theory, the original
gauge potential $A_{\mu}$ has been replaced by a massive spin--1 field
$v_{\mu}$. Now, observe that due to (\ref{CC11}) the functional
derivative with respect to $\xi^{\mu}$ produces the term $-\alpha_j e
v_\mu$. Therefore, the non--linear (and non--local) term $\left( k^{\mu
\bot}[\tfrac{\delta}{{\rm i} \,
\delta \xi^{\mu}},|h|] \,k_{\mu}^{\bot}[\tfrac{\delta}{{\rm i} \,
\delta \xi^{\mu}},|h|]\right)$ is effectively of the order $e^4$, and
we can treat it as a perturbation. Performing the functional derivatives
in (\ref{CC10}) with respect to $\xi^{\mu}$ yields a complicated non--local 
measure, which in full generality cannot be handled analytically. 
Limiting ourselves to lowest order we get:
\begin{eqnarray}
\lefteqn{
<0|T j^{\mu}(y_1) \, j^{\nu}(y_2)|0>_0
}
\nonumber \\
 & = & \frac{1}{Z[0]} \,
       \int \prod_x \big\{ {\rm d} v_{\mu} \, {\rm d} |h|^2 \big\} \nonumber \\
 &   & \hspace*{1cm} \times \,
       \left( \alpha_j \, \eta^{\mu \nu} \, \delta^4(y_1-y_2)
            + m_v^2 \alpha_j \, v^{\mu}(y_1) v^{\nu}(y_2)
       \right) \,
       e^{{\rm i} \int {\rm d}^4 x \, {\cal L}_0[\eta^{\mu};v_{\mu},|h|]}
       \big|_{\eta^{\mu} = 0}. \nonumber \\
 \label{CC12}
\end{eqnarray}
Here we have denoted
\begin{equation*}
Z[0] = \int \prod_x \big\{ {\rm d} v_{\mu} \, {\rm d} |h|^2 \big\} \,
       e^{{\rm i} \int {\rm d}^4 x \, {\cal L}_0[0;v_{\mu},|h|]}
\end{equation*}
and
\begin{eqnarray}
{\cal L}_0[\eta^{\mu};v_{\mu},|h|]   =
   - \tfrac{1}{4} \, (\partial_{[\mu} v_{\nu]})^2
   + \tfrac{m_v^2}{2} \, v^{\mu} v_{\mu}
   - \tfrac{1}{2 \, \alpha_h} |h|^4
   + \eta^{\mu} v_{\mu}\, . \label{CC13}
\end{eqnarray}
Next, observe that the integration over $|h|$ decouples, giving
\begin{eqnarray}
\lefteqn{ <0|T j^{\mu}(y_1) \, j^{\nu}(y_2)|0>_0 } \nonumber \\
 &  = &
       \frac{1}{Z[0]} \, \int \prod_x \big\{ {\rm d} v_{\mu} \big\}
       \left( \alpha_j \, \eta^{\mu \nu} \, \delta^4(y_1-y_2)
            + m_v^2 \alpha_j \, v^{\mu}(y_1) v^{\nu}(y_2)
       \right) \,
       e^{{\rm i} \int {\rm d}^4 x \, {\cal L}_0[\eta^{\mu};v_{\mu}]}
       \big|_{\eta^{\mu} = 0} \nonumber \\
	& \equiv &
       <\alpha_j \, \eta^{\mu \nu} \, \delta^4(y_1-y_2)
        + m_v^2 \alpha_j \, v^{\mu}(y_1) v^{\nu}(y_2)> \ ,
\nonumber
\end{eqnarray}
where now
$$Z[0] = \int \prod_x \big\{ {\rm d} v_{\mu} \big\} \,
       e^{{\rm i} \int {\rm d}^4 x \, {\cal L}_0[0;v_{\mu}]}$$
and
$${\cal L}_0[\eta^{\mu};v_{\mu}]
 = - \tfrac{1}{4} \, (\partial_{[\mu} v_{\nu]})^2
   + \tfrac{m_v^2}{2} \, v^{\mu} v_{\mu}
   + \eta^{\mu} v_{\mu}\, .$$

This way we obtain:
\begin{proposition}
In lowest order, the current--current propagator $<0|T j^{\mu}(x) \, j^{\nu}(y)|0>$ of
massless QED is given by the vacuum expectation value
\begin{equation}
\label{CCRes1}
<0|T j^{\mu}(y_1) \, j^{\nu}(y_2)|0>_0
 = <\alpha_j \, \eta^{\mu \nu} \, \delta^4(y_1-y_2)
    + m_v^2 \alpha_j \, v^{\mu}(y_1) v^{\nu}(y_2)>
\end{equation}
with respect to the functional measure
$\prod_x \big\{ {\rm d} v_{\mu} \big\} \,
 e^{{\rm i} \int {\rm d}^4 x \, {\cal L}[v_{\mu}]}$
and the Lagrangian
\begin{equation}
{\cal L}[v_{\mu}]
 = - \tfrac{1}{4} \, (\partial_{[\mu} v_{\nu]})^2
   + \tfrac{m_v^2}{2} \, v^{\mu} v_{\mu} \ .
\end{equation}
\end{proposition}

Thus, in lowest order, we are left with a simple Gaussian integration.
We have
$${\cal L}_0[\eta^{\mu};v_{\mu}]
 = - \tfrac{1}{4} \, (\partial_{[\mu} v_{\nu]})^2
   + \tfrac{m_v^2}{2} \, v^{\mu} v_{\mu}
   + \eta^{\mu} v_{\mu}
 = - \tfrac{1}{2} \, v_{\mu} {D^{\mu}}_{\nu} v^{\nu}
   + \eta^{\mu} v_{\mu}\  ,$$
with
${D^{\nu}}_{\mu}
    := \partial^{\nu} \partial_{\mu}
     - \delta^{\nu}_{\mu} \partial^{\sigma} \partial_{\sigma}
     - m_v^2 \, \delta^{\nu}_{\mu}$.
We introduce the new field
\begin{equation*}
v^{\mu'}(x)
  := v^{\mu}(x) + \int {\rm d}^4y \eta^{\nu}(y)
  {(D^{-1})^{\mu}}_{\nu}(x-y) \ ,
\end{equation*}
where the propagator ${(D^{-1})^{\mu}}_{\nu}(x-y)$ of the massive free
field $v_{\mu}$ is defined by
\begin{equation*}
{D^{\mu}}_{\alpha} \, {(D^{-1})^{\alpha}}_{\nu}(x-y)
  = - \, \delta^4(x-y) \, \delta^{\mu}_{\nu}\  .
\end{equation*}
Thus,
\begin{equation}
\label{CC15}
{(D^{-1})^{\mu}}_{\nu}(x-y)
 = \tfrac{\partial^{\mu} \partial_{\nu} + m_v^2 \, \delta^{\mu}_{\nu}}
        {m_v^2 \, (m_v^2 + \Box)} \,
   \delta^4(x-y)  \  .
\end{equation}
Performing the above transformation, the resulting integration over
$v_{\mu}'$ decouples and we obtain
\begin{eqnarray}
\lefteqn{ <0|T j^{\mu}(y_1) \, j^{\nu}(y_2)|0>_0 } \nonumber \\
 &\hspace*{1cm} = &
       \left( \alpha_j \, \eta^{\mu \nu} \, \delta^4(y_1-y_2)
            + m_v^2 \alpha_j \, \frac{1}{{\rm i}^2} \, \frac{\delta}{\delta \eta_{\mu}(y_1)} \,
              \frac{\delta}{\delta \eta_{\nu}(y_2)}
       \right)
       \nonumber \\  &   & \times \,
       e^{{\rm i} \int {\rm d}^4 x {\rm d}^4y \, \big\{
         - \eta_{\nu}(x) \, {(D^{-1})^{\nu}}_{\mu}(x-y) \, \eta^{\mu}(y)
         \big\} }
       \big|_{\eta^{\mu} = 0}. \label{CC14}
\end{eqnarray}
Finally, performing the remaining functional derivatives, we get
\begin{eqnarray}
\label{CCRes2}
<0|T j^{\mu}(y_1) \, j^{\nu}(y_2)|0>_0
 & = & \alpha_j \, \eta^{\mu \nu} \, \delta^4(y_1-y_2)
     - \alpha_j \,
       \tfrac{\partial^{\mu} \partial^{\nu} + m_v^2 \, \eta^{\mu \nu}}
             {m_v^2 + \Box} \,
       \delta^4(y_1-y_2) \nonumber \\
 & = & \alpha_j \, \tfrac{\eta^{\mu \nu} \, \Box
                          - \partial^{\mu} \partial^{\nu}}{m_v^2 + \Box} \,
       \delta^4(y_1-y_2).
\end{eqnarray}
Fourier transforming to momentum space leads to the following expression
in lowest order:
\begin{equation}
{\Pi}_0^{\mu \nu}(p)
 = {\cal F} <0|T j^{\mu}(y_1) \, j^{\nu}(y_2)|0>_0
 = (p^{\mu} p^{\nu} - \eta^{\mu \nu} p^2) \, T(p^2)
   \label{QEDCCfour}
\end{equation}
with $T(p^2) := \frac{\alpha_j}{m_v^2 - p^2}$. This result has the
expected Lorentz structure. Moreover, we obtain the identity $p_{\mu} \,
{\Pi}_0^{\mu \nu}(p) = 0$, which is nothing but the vector Ward
identity. Thus, in lowest order of the above defined perturbation series,
our result obeys the vector Ward identity.

However, a direct comparison with the well--known perturbation series of
QED is not possible. This is due to the fact that the mass $m_v^2 =
\alpha_j e^2$ of the spin--1 field $v_{\mu}$ occurring in (\ref{CCRes2}) 
contains the bare coupling
constant $e$. Thus, expanding this result around $m_v^2 = 0$, which
corresponds to a power expansion in $e^2$, we obtain non--vanishing
contributions to all orders in $e^2$. Therefore, formula (\ref{CCRes2})
can be interpreted as a resumation of particular quantum corrections,
which results in an effective (``dynamical'') mass for the field
$v_{\mu}$. Unfortunately, higher order contributions (in the sense of
our expansion) cannot be calculated analytically.

\section{The Chiral Anomaly}
\label{ChiralAnomaly}

In this section we show that within our approach the correct
Adler-Bardeen anomaly \cite{AB} is obtained in the lowest order
approximation discussed in the previous section. Again we restrict
ourselves to the massless case.

Due to the bosonization rule (\ref{QEDBos}) the chiral anomaly
$<\partial_{\mu} j^{\mu}_5>$ is given by the vacuum expectation value
\begin{eqnarray}
\label{CA1}
<\partial_{\mu} j^{\mu}_5>
 & \equiv & <\tfrac{\alpha_{j^5}}{2} \, (\partial_{\mu}
 \partial^{\mu} \alpha(y))>
       \nonumber \\
 & = & \frac{1}{Z[0,0,0]} \, \frac{\alpha_{j^5}}{2 {\rm i}} \,
       \left( \partial_{\mu} \, \frac{\delta}{\delta \zeta_{\mu}(y)}
       \right) \,
       Z[\zeta^{\mu}, \xi^{\mu}, \eta^{\mu}]
       \Big|_{\zeta^{\mu},\xi^{\mu}, \eta^{\mu} = 0}  \  ,
\end{eqnarray}
where $Z[\zeta^{\mu}, \xi^{\mu}, \eta^{\mu}]$ is given by (\ref{GFI}) and
${\cal L}[\zeta^{\mu}, \xi^{\mu}, \eta^{\mu}; v_{\mu},j^{\mu},|h|,\alpha]$
by (\ref{GLag}), or in terms of $k^{\mu}[j_{\mu},|h|]$, by (\ref{CC5}).
Choosing a longitudinal source current
$\zeta^{\mu} := (\partial^{\mu} \zeta)$
and transforming
\begin{equation}
\label{CA4}
\alpha''
 = \alpha + \tfrac{4}{\alpha_{j^5}} \, \zeta
 + \tfrac{1}{4 \, \alpha_{j^5}} \, k[j_{\mu},|h|]
\end{equation}
yields the Lagrangian in the following form:
\begin{eqnarray}
\label{CA5}
\lefteqn{ {\cal L}[\zeta^{\mu}, \xi^{\mu}, \eta^{\mu};
v_{\mu},j^{\mu},|h|,\alpha'']}
\nonumber \\
& \hspace*{1cm} = &
    - \tfrac{1}{4} (\partial_{[\mu}\,v_{\nu]})^2
    - e \, j^{\mu} v_{\mu}
    - \tfrac{1}{4 \, \alpha_{j^5}} \, \zeta_{\mu} k^{\mu \|}[j_{\mu},|h|]
    + \tfrac{\alpha_{j^5}}{128} \, k^{\mu \bot}[j_{\mu},|h|]
      k_{\mu}^{\bot}[j_{\mu},|h|]
    \nonumber \\
& & - \tfrac{1}{2 \, \alpha_j} \, j^{\mu} j_{\mu}
    - \tfrac{1}{2 \, \alpha_h} |h|^4
    - \tfrac{2}{\alpha_{j^5}} \, \zeta_{\mu} \zeta^{\mu}
    + \tfrac{\alpha_{j^5}}{8} \, (\partial^{\mu} \alpha'')
      (\partial_{\mu} \alpha'')
    + \xi^{\mu} j_{\mu}
    + \eta^{\mu} v_{\mu}. \nonumber
\end{eqnarray}
Now $\alpha''$ can be integrated out trivially. Performing the
functional derivative with respect to $\zeta^{\mu}$ we obtain
\begin{eqnarray*}
\lefteqn{ <\tfrac{\alpha_{j^5}}{2} \,
           (\partial_{\mu} \partial^{\mu} \alpha(y))> } \\
 & \hspace*{1cm} = & \frac{1}{Z[0,0,0]} \,
       \int \prod_x \big\{ {\rm d} v_{\mu} \, {\rm d} j^{\mu} \,
       {\rm d} |h|^2 \big\} \,
       \left( - \, \tfrac{1}{8} \, \partial_{\mu}
              (\partial^{\mu}k[j_{\mu}(y),|h(y)|] + 4 \,
              \zeta^{\mu}(y)) \right) \\
 &   & \hspace*{1.7cm} \times \,
       e^{{\rm i} \int {\rm d}^4 x \,
         {\cal L}[\zeta^{\mu}, \xi^{\mu}, 0; v_{\mu},j^{\mu},|h|]}
       \Big|_{\zeta^{\mu}, \xi^{\mu} = 0} \\
 & \hspace*{1cm} = & \frac{1}{Z[0,0,0]} \,
       \int \prod_x \big\{ {\rm d} v_{\mu} \, {\rm d} j^{\mu} \,
       {\rm d} |h|^2 \big\} \,
       \left( - \, \tfrac{1}{8} \, \partial_{\mu}
              k^{\mu}[j_{\mu}(y),|h(y)|]
       \right) \\
 &   & \hspace*{1.7cm} \times \,
       e^{{\rm i} \int {\rm d}^4 x \,
         {\cal L}[0, \xi^{\mu}, 0; v_{\mu},j^{\mu},|h|]}
       \Big|_{\xi^{\mu} = 0},
\end{eqnarray*}
where
\begin{eqnarray*}
\lefteqn{ {\cal L}[0, \xi^{\mu}, 0; v_{\mu},j^{\mu},|h|] } \\
& \hspace*{0.5cm} = &
      - \tfrac{1}{4} \, (\partial_{[\mu} v_{\nu]})^2
      - e \, j^{\mu} v_{\mu}
      + \tfrac{\alpha_{j^5}}{128} \, k^{\mu \bot}[j_{\mu},|h|]
        k_{\mu}^{\bot}[j_{\mu},|h|]
      - \tfrac{1}{2 \, \alpha_j} \, j^{\mu} j_{\mu}
      - \tfrac{1}{2 \, \alpha_h} |h|^4
      + \xi^{\mu} j_{\mu}.
\end{eqnarray*}

Due to the explicit dependence of $k^{\nu}[j_{\mu}(y),|h(y)|]$ on $|h|$,
a further exact analytical treatment of this formula is impossible. But,
as outlined in Section \ref{SEffectiveBosonizedForm}, $|h|$ enters our
effective theory in a non--dynamical way and, therefore, in principle it can
be ``averaged'' out. In the simplest approximation we replace $|h|$ by
its mean value $\chi_0$. This way we are led to
\begin{eqnarray}
\lefteqn{ <\tfrac{\alpha_{j^5}}{2} \,
           (\partial_{\mu} \partial^{\mu} \alpha(y))> } \nonumber \\
 & \hspace*{1cm} = &
       \frac{1}{Z[0,0,0]} \,
       \int \prod_x \big\{ {\rm d} v_{\mu} \, {\rm d} j^{\mu} \big\} \,
       \left( - \, \tfrac{1}{8} \, \partial_{\mu}
       k^{\mu}[j_{\mu}(y),\chi_0] \right)
       \nonumber \\
 &   & \hspace*{1.7cm} \times \,
       e^{{\rm i} \int {\rm d}^4 x \,
         {\cal L}[0, \xi^{\mu}, 0; v_{\mu},j^{\mu}]}
       \Big|_{\xi^{\mu} = 0} \nonumber \\
 & \hspace*{1cm} = & \frac{1}{Z[0,0,0]} \,
    \int \prod_x \big\{ {\rm d} v_{\mu} \, {\rm d} j^{\mu} \big\}
     \left( - \tfrac{1}{8 \, \chi_0^2} \,
             \epsilon^{\delta \mu \sigma \xi}
             (\partial_{\mu} j_{\delta}(y))
             (\partial_{\sigma} j_{\xi}(y)) \right) \nonumber \\ &   &
 \hspace*{1.7cm} \times \,
    \, e^{{\rm i} \int {\rm d}^4 x \,
      {\cal L}[0, \xi^{\mu}, 0; v_{\mu},j^{\mu}]}
    \Big|_{\xi^{\mu} = 0}, \label{CA6}
\end{eqnarray}
with
\begin{eqnarray*}
\lefteqn{ {\cal L}[0, \xi^{\mu}, 0; v_{\mu},j^{\mu}] } \\
 & \hspace*{1cm} = &
   - \tfrac{1}{4} \, (\partial_{[\mu} v_{\nu]})^2
   - e \, j^{\mu} v_{\mu}
   + \tfrac{\alpha_{j^5}}{128} \, k^{\mu \bot}[j_{\mu},\chi_0]
     k_{\mu}^{\bot}[j_{\mu},\chi_0]
   - \tfrac{1}{2 \, \alpha_j} \, j^{\mu} j_{\mu}
   + \xi^{\mu} j_{\mu}.
\end{eqnarray*}

Now the non--trivial coupling term $\frac{\alpha_{j^5}}{128} \,
 k^{\mu \bot}[j_{\mu},\chi_0] \, k_{\mu}^{\bot}[j_{\mu},\chi_0]$ will be
treated similarly as in the previous Section. In lowest order we get
\begin{eqnarray*}
\lefteqn{ <\tfrac{\alpha_{j^5}}{2} \,
           (\partial_{\mu} \partial^{\mu} \alpha(y))>_0 } \\ &
 = & \frac{1}{Z[0,0,0]} \,
       \int \prod_x \Big\{ {\rm d} v_{\mu} \, {\rm d} j^{\mu} \Big\} \,
          \left( - \tfrac{1}{8 \, \chi_0^2} \,
             \epsilon^{\delta \mu \sigma \xi}
             (\partial_{\mu} j_{\delta}(y))
             (\partial_{\sigma} j_{\xi}(y))
    \right) \,
    e^{{\rm i} \int {\rm d}^4 x \,
      {\cal L}_0[0, \xi^{\mu}, 0; v_{\mu},j^{\mu}]}
    \Big|_{\xi^{\mu} = 0} \\
 & = & \frac{1}{Z[0,0,0]} \,
       \int \prod_x \Big\{ {\rm d} v_{\mu} \, {\rm d} j^{\mu} \Big\} \,
             \left( - \tfrac{1}{8 {\rm i}^2 \, \chi_0^2} \,
                \epsilon^{\delta \mu \sigma \xi} \big( \partial_{\mu}
 \frac{\delta}{\delta \xi^{\delta}(y)} \big) \big( \partial_{\sigma}
 \frac{\delta}{\delta \xi^{\xi}(y)} \big) \right) \\
 &   &
 \hspace*{1.7cm} \times \,
       e^{{\rm i} \int {\rm d}^4 x \,
          {\cal L}_0[0, \xi^{\mu}, 0; v_{\mu},j^{\mu}]}
       \Big|_{\xi^{\mu} = 0},
\end{eqnarray*}
where
\begin{eqnarray*}
Z[0,0,0]  =
       \int \prod_x \Big\{ {\rm d} v_{\mu} \, {\rm d} j^{\mu} \Big\} \,
         e^{{\rm i} \int {\rm d}^4 x \,
         {\cal L}_0[0,0,0; v_{\mu},j^{\mu}]}
     \end{eqnarray*}
and
\begin{equation*}
{\cal L}_0[0, \xi^{\mu}, 0; v_{\mu},j^{\mu}]
 = - \tfrac{1}{4} \, (\partial_{[\mu} v_{\nu]})^2
   - e \, j^{\mu} v_{\mu}
   - \tfrac{1}{2 \, \alpha_j} \, j^{\mu} j_{\mu}
   + \xi^{\mu} j_{\mu}.
\end{equation*}

Performing the Gaussian integration over $j^{\mu}$ and the functional
derivatives with respect to $\xi^{\mu}$, we finally obtain
\begin{eqnarray}
\lefteqn{ <\tfrac{\alpha_{j^5}}{2} \,
           (\partial_{\mu} \partial^{\mu} \alpha(y))>_0 } \nonumber \\
 & \hspace*{1cm} = & \tfrac{e^2 \alpha_j^2}{8 \chi_0^2} \,
       \epsilon^{\mu \nu \alpha \beta} \,
       \frac{1}{Z[0]} \,
       \int \prod_x \big\{ {\rm d} v_{\mu} \big\} \,
       (\partial_{[\mu} v_{\nu]}(y)) (\partial_{[\alpha} v_{\beta]}(y)) \,
       e^{{\rm i} \int {\rm d}^4 x \, {\cal L}[v^{\mu}]} \nonumber \\
 & \hspace*{1cm} = & \tfrac{e^2 \alpha_j^2}{8 \chi_0^2} \,
       \epsilon^{\mu \nu \alpha \beta} \,
       <(\partial_{[\mu} v_{\nu]}(y)) (\partial_{[\alpha} v_{\beta]}(y))>
       \label{CA7}
\end{eqnarray}
with
$Z[0] = \int \prod_x \big\{ {\rm d} v_{\mu} \big\} \,
        e^{{\rm i} \int {\rm d}^4 x \, {\cal L}[v^{\mu}]}$
and
\begin{equation}
{\cal L}[v_{\mu}]
 = - \tfrac{1}{4} \, (\partial_{[\mu} v_{\nu]})^2
   + \tfrac{m_v^2}{2} \, v^{\mu} v_{\mu} \ .
\end{equation}

Thus, bearing in mind that $F_{\mu \nu} = \partial_{[\mu} v_{\nu]}$
we can formulate the following result:

\begin{proposition}
In lowest order, the chiral anomaly of massless QED in $(3+1)$ dimensions
is given by the vacuum expectation value
\begin{equation}
\label{CARes}
<\tfrac{\alpha_{j^5}}{2} \, (\partial_{\mu} \partial^{\mu} \alpha(y))>_0
 = \tfrac{e^2 \alpha_j^2}{8 \chi_0^2} \,
   \epsilon^{\mu \nu \alpha \beta} \,
   <F_{\mu \nu} \, F_{\alpha \beta}>
\end{equation}
with respect to the functional measure $\prod_x \big\{ {\rm d} v_{\mu}
\big\} \,
 e^{{\rm i} \int {\rm d}^4 x \, {\cal L}[v_{\mu}]}$ \  ,
where
\begin{equation}
{\cal L}[v_{\mu}]
 = - \tfrac{1}{4} \, (\partial_{[\mu} v_{\nu]})^2
   + \tfrac{m_v^2}{2} \, v^{\mu} v_{\mu}  \  .
\end{equation}
\end{proposition}

If we start with an external electromagnetic field, $v_{\mu}$
becomes external, too. In that case we get
\begin{equation}
\label{CARes2}
<\tfrac{\alpha_{j^5}}{2} \, (\partial_{\mu} \partial^{\mu} \alpha(y))>_v
 = \tfrac{e^2 \alpha_j^2}{8 \chi_0^2} \,
   \epsilon^{\mu \nu \alpha \beta} \, F_{\mu \nu} \, F_{\alpha \beta},
\end{equation}
where $F_{\mu \nu} = \partial_{[\mu} v_{\nu]}$ denotes the external
electromagnetic field strength.

Observe that we get the correct coefficient, see \cite{Bertl}, \cite{AB},
for the chiral anomaly if we require the following relation between our
parameters:
\begin{equation}
\label{CAFix}
\tfrac{\alpha_j^2}{\chi_0^2} \, = \tfrac{1}{2 \, \pi^2} \  .
\end{equation}
This yields
\begin{equation}
\label{CARes3}
<\partial^{\mu} j_{\mu}^5(y))>_v
 \equiv <\tfrac{\alpha_{j^5}}{2} \, (\partial_{\mu} \partial^{\mu}
 \alpha(y))>_v
   = \tfrac{e^2}{16 \, \pi^2} \,
     \epsilon^{\mu \nu \alpha \beta} \, F_{\mu \nu} \, F_{\alpha \beta}.
\end{equation}
Thus, the anomaly can be used to fix one of the parameters of our
effective theory as discussed in the Introduction.


\end{document}